\documentclass[tickz,longauth]{aa}
\usepackage{tikz}

\usepackage{amsmath}
\usepackage{natbib, twoopt}

\usepackage{graphicx}
\usepackage[varg]{txfonts}
\usepackage{color}
\usepackage{url}
\usepackage{wasysym}

\usepackage{soul}
\sethlcolor{yellow}
\usepackage[normalem]{ulem}

\usepackage{hyperref}

\usepackage[switch, modulo]{lineno}
\setcitestyle{citesep={,}}

\usepackage{booktabs} 
\usepackage{siunitx}

\usepackage{subcaption}
\usepackage{shortvrb}
\MakeShortVerb{\|}
\usepackage{txfonts}

\makeatletter
\newcommand\tinyv{\@setfontsize\tinyv{4pt}{6}}
\renewcommand*{\@fnsymbol}[1]{\ifcase#1\or*\or$\dagger$\or$\ddagger$\or**\or$\dagger\dagger$\or$\ddagger\ddagger$\fi}
\makeatother

\newcommand{\gray}{$\gamma$-ray}
\newcommand{\grays}{$\gamma$ rays}

\def\araa{ARA\&A}             
\def\apj{ApJ}                 
\def\apjl{ApJ}                
\def\apjs{ApJS}               
\def\aap{A\&A}                
\def\aapr{A\&A~Rev.}          
\def\mnras{MNRAS}             
\def\pasj{PASJ}               

\def\nat{Nature}              

\makeatletter
\renewcommand*{\@fnsymbol}[1]{\ifcase#1\or*\or$\dagger$\or$\ddagger$\or**\or$\dagger\dagger$\or$\ddagger\ddagger$\fi}
\makeatother

\begin{document}
\title{Evidence for $\gamma$-ray emission from the remnant of Kepler's supernova based on deep H.E.S.S. observations}

\authorrunning{H.E.S.S. Collaboration}
\titlerunning{Evidence for $\gamma$ rays from Kepler's SNR}

\author{\fontsize{11.0}{13.0}\selectfont{H.E.S.S. Collaboration
\and F.~Aharonian \inst{\ref{DIAS},\ref{MPIK},\ref{RAU}}
\and F.~Ait~Benkhali \inst{\ref{LSW}}
\and E.O.~Ang\"uner \inst{\ref{CPPM}}
\and H.~Ashkar \inst{\ref{LLR}}
\and M.~Backes \inst{\ref{UNAM},\ref{NWU}}
\and V.~Barbosa~Martins \inst{\ref{DESY}}
\and R.~Batzofin \inst{\ref{Wits}}
\and Y.~Becherini \inst{\ref{APC},\ref{Linnaeus}}
\and D.~Berge \inst{\ref{DESY}}
\and K.~Bernl\"ohr \inst{\ref{MPIK}}
\and M.~B\"ottcher \inst{\ref{NWU}}
\and C.~Boisson \inst{\ref{LUTH}}
\and J.~Bolmont \inst{\ref{LPNHE}}
\and M.~de~Bony~de~Lavergne \inst{\ref{LAPP}}
\and M.~Breuhaus \inst{\ref{MPIK}}
\and R.~Brose \inst{\ref{DIAS}}
\and F.~Brun \inst{\ref{CEA}}
\and T.~Bulik \inst{\ref{UWarsaw}}
\and T.~Bylund \inst{\ref{Linnaeus}}
\and F.~Cangemi \inst{\ref{LPNHE}}
\and S.~Caroff \inst{\ref{LPNHE}}
\and S.~Casanova \inst{\ref{IFJPAN}}
\and M.~Cerruti \inst{\ref{APC}}
\and T.~Chand \inst{\ref{NWU}}
\and A.~Chen \inst{\ref{Wits}}
\and O.~Chibueze \inst{\ref{NWU}}
\and G.~Cotter \inst{\ref{Oxford}}
\and P.~Cristofari \inst{\ref{LUTH}}
\and J.~Damascene~Mbarubucyeye \inst{\ref{DESY}}
\and J.~Devin \inst{\ref{CENBG}}
\and A.~Djannati-Ata\"i \inst{\ref{APC}}
\and A.~Dmytriiev \inst{\ref{LUTH}}
\and K.~Egberts \inst{\ref{UP}}
\and S.~Einecke \inst{\ref{Adelaide}}
\and J.-P.~Ernenwein \inst{\ref{CPPM}}
\and K.~Feijen \inst{\ref{Adelaide}}
\and A.~Fiasson \inst{\ref{LAPP}}
\and G.~Fichet~de~Clairfontaine \inst{\ref{LUTH}}
\and G.~Fontaine \inst{\ref{LLR}}
\and S.~Funk \inst{\ref{ECAP}}
\and S.~Gabici \inst{\ref{APC}}
\and Y.A.~Gallant \inst{\ref{LUPM}}
\and S.~Ghafourizadeh \inst{\ref{LSW}}
\and G.~Giavitto \inst{\ref{DESY}}
\and L.~Giunti \inst{\ref{APC},\ref{CEA}}
\and D.~Glawion \inst{\ref{ECAP}}
\and J.F.~Glicenstein \inst{\ref{CEA}}
\and M.-H.~Grondin \inst{\ref{CENBG}}
\and M.~H\"{o}rbe \inst{\ref{Oxford}}
\and W.~Hofmann \inst{\ref{MPIK}}
\and T.~L.~Holch \inst{\ref{DESY}}
\and M.~Holler \inst{\ref{Innsbruck}}
\and D.~Horns \inst{\ref{UHH}}
\and Zhiqiu~Huang \inst{\ref{MPIK}}
\and M.~Jamrozy \inst{\ref{UJK}}
\and V.~Joshi \inst{\ref{ECAP}}
\and I.~Jung-Richardt \inst{\ref{ECAP}}
\and E.~Kasai \inst{\ref{UNAM}}
\and K.~Katarzy{\'n}ski \inst{\ref{NCUT}}
\and U.~Katz \inst{\ref{ECAP}}
\and B.~Kh\'elifi \inst{\ref{APC}}
\and W.~Klu\'{z}niak \inst{\ref{NCAC}}
\and Nu.~Komin \inst{\ref{Wits}}
\and K.~Kosack \inst{\ref{CEA}}
\and D.~Kostunin \inst{\ref{DESY}}
\and A.~Lemi\`ere \inst{\ref{APC}}
\and M.~Lemoine-Goumard \inst{\ref{CENBG}}
\and J.-P.~Lenain \inst{\ref{LPNHE}}
\and F.~Leuschner \inst{\ref{IAAT}}
\and T.~Lohse \inst{\ref{HUB}}
\and A.~Luashvili \inst{\ref{LUTH}}
\and I.~Lypova \inst{\ref{LSW}}
\and J.~Mackey \inst{\ref{DIAS}}
\and D.~Malyshev \inst{\ref{IAAT}}
\and D.~Malyshev \inst{\ref{ECAP}}
\and V.~Marandon \inst{\ref{MPIK}}
\and P.~Marchegiani \inst{\ref{Wits}}
\and A.~Marcowith \inst{\ref{LUPM}}
\and G.~Mart\'i-Devesa \inst{\ref{Innsbruck}}
\and R.~Marx \inst{\ref{LSW}}
\and G.~Maurin \inst{\ref{LAPP}}
\and P.J.~Meintjes \inst{\ref{UFS}}
\and M.~Meyer \inst{\ref{UHH}}
\and A.~Mitchell \inst{\ref{ECAP},\ref{MPIK}}
\and R.~Moderski \inst{\ref{NCAC}}
\and L.~Mohrmann \inst{\ref{MPIK}}
\and A.~Montanari \inst{\ref{CEA}}
\and E.~Moulin \inst{\ref{CEA}}
\and J.~Muller \inst{\ref{LLR}}
\and K.~Nakashima \inst{\ref{ECAP}}
\and M.~de~Naurois \inst{\ref{LLR}}
\and A.~Nayerhoda \inst{\ref{IFJPAN}}
\and J.~Niemiec \inst{\ref{IFJPAN}}
\and A.~Priyana~Noel \inst{\ref{UJK}}
\and P.~O'Brien \inst{\ref{Leicester}}
\and S.~Ohm \inst{\ref{DESY}}
\and L.~Olivera-Nieto \inst{\ref{MPIK}}
\and E.~de~Ona~Wilhelmi \inst{\ref{DESY}}
\and M.~Ostrowski \inst{\ref{UJK}}
\and S.~Panny \inst{\ref{Innsbruck}}
\and M.~Panter \inst{\ref{MPIK}}
\and R.D.~Parsons \inst{\ref{HUB}}
\and G.~Peron \inst{\ref{MPIK}}
\and V.~Poireau \inst{\ref{LAPP}}
\and D.A.~Prokhorov \inst{\ref{Amsterdam}}\protect\footnotemark[1]
\and G.~P\"uhlhofer \inst{\ref{IAAT}}
\and M.~Punch \inst{\ref{APC},\ref{Linnaeus}}
\and A.~Quirrenbach \inst{\ref{LSW}}
\and P.~Reichherzer \inst{\ref{CEA}}
\and A.~Reimer \inst{\ref{Innsbruck}}
\and O.~Reimer \inst{\ref{Innsbruck}}
\and M.~Renaud \inst{\ref{LUPM}}
\and B.~Reville \inst{\ref{MPIK}}
\and F.~Rieger \inst{\ref{MPIK}}
\and G.~Rowell \inst{\ref{Adelaide}}
\and B.~Rudak \inst{\ref{NCAC}}
\and H.~Rueda Ricarte \inst{\ref{CEA}}
\and V.~Sahakian \inst{\ref{Yerevan}}
\and S.~Sailer \inst{\ref{MPIK}}
\and H.~Salzmann \inst{\ref{IAAT}}
\and D.A.~Sanchez \inst{\ref{LAPP}}
\and A.~Santangelo \inst{\ref{IAAT}}
\and M.~Sasaki \inst{\ref{ECAP}}
\and J.~Sch\"afer \inst{\ref{ECAP}}
\and F.~Sch\"ussler \inst{\ref{CEA}}
\and H.M.~Schutte \inst{\ref{NWU}}
\and U.~Schwanke \inst{\ref{HUB}}
\and J.N.S.~Shapopi \inst{\ref{UNAM}}
\and R.~Simoni \inst{\ref{Amsterdam}}\protect\footnotemark[1]
\and H.~Sol \inst{\ref{LUTH}}
\and A.~Specovius \inst{\ref{ECAP}}
\and S.~Spencer \inst{\ref{Oxford}}
\and {\L.}~Stawarz \inst{\ref{UJK}}
\and S.~Steinmassl \inst{\ref{MPIK}}
\and C.~Steppa \inst{\ref{UP}}
\and I.~Sushch \inst{\ref{NWU}}
\and T.~Takahashi \inst{\ref{KAVLI}}
\and T.~Tanaka \inst{\ref{Konan}}
\and A.M.~Taylor \inst{\ref{DESY}}
\and R.~Terrier \inst{\ref{APC}}
\and M.~Tsirou \inst{\ref{MPIK}}
\and Y.~Uchiyama \inst{\ref{Rikkyo}}
\and T.~Unbehaun \inst{\ref{ECAP}}
\and C.~van~Eldik \inst{\ref{ECAP}}
\and J.~Veh \inst{\ref{ECAP}}
\and J.~Vink \inst{\ref{Amsterdam}}\protect\footnotemark[1]
\and H.J.~V\"olk \inst{\ref{MPIK}}
\and S.J.~Wagner \inst{\ref{LSW}}
\and F.~Werner \inst{\ref{MPIK}}
\and R.~White \inst{\ref{MPIK}}
\and A.~Wierzcholska \inst{\ref{IFJPAN}}
\and Yu~Wun~Wong \inst{\ref{ECAP}}
\and A.~Yusafzai \inst{\ref{ECAP}}
\and M.~Zacharias \inst{\ref{LUTH},\ref{NWU}}
\and D.~Zargaryan \inst{\ref{DIAS},\ref{RAU}}
\and A.A.~Zdziarski \inst{\ref{NCAC}}
\and A.~Zech \inst{\ref{LUTH}}
\and S.J.~Zhu \inst{\ref{DESY}}
\and S.~Zouari \inst{\ref{APC}}
\and N.~\.Zywucka \inst{\ref{NWU}}
}}

\institute{
Dublin Institute for Advanced Studies, 31 Fitzwilliam Place, Dublin 2, Ireland \label{DIAS} \and
Max-Planck-Institut f\"ur Kernphysik, P.O. Box 103980, D 69029 Heidelberg, Germany \label{MPIK} \and
High Energy Astrophysics Laboratory, RAU,  123 Hovsep Emin St  Yerevan 0051, Armenia \label{RAU} \and
Landessternwarte, Universit\"at Heidelberg, K\"onigstuhl, D 69117 Heidelberg, Germany \label{LSW} \and
Aix Marseille Universit\'e, CNRS/IN2P3, CPPM, Marseille, France \label{CPPM} \and
Laboratoire Leprince-Ringuet, École Polytechnique, CNRS, Institut Polytechnique de Paris, F-91128 Palaiseau, France \label{LLR} \and
University of Namibia, Department of Physics, Private Bag 13301, Windhoek 10005, Namibia \label{UNAM} \and
Centre for Space Research, North-West University, Potchefstroom 2520, South Africa \label{NWU} \and
DESY, D-15738 Zeuthen, Germany \label{DESY} \and
School of Physics, University of the Witwatersrand, 1 Jan Smuts Avenue, Braamfontein, Johannesburg, 2050 South Africa \label{Wits} \and
Université de Paris, CNRS, Astroparticule et Cosmologie, F-75013 Paris, France \label{APC} \and
Department of Physics and Electrical Engineering, Linnaeus University,  351 95 V\"axj\"o, Sweden \label{Linnaeus} \and
Laboratoire Univers et Théories, Observatoire de Paris, Université PSL, CNRS, Université de Paris, 92190 Meudon, France \label{LUTH} \and
Sorbonne Universit\'e, Universit\'e Paris Diderot, Sorbonne Paris Cit\'e, CNRS/IN2P3, Laboratoire de Physique Nucl\'eaire et de Hautes Energies, LPNHE, 4 Place Jussieu, F-75252 Paris, France \label{LPNHE} \and
Université Savoie Mont Blanc, CNRS, Laboratoire d'Annecy de Physique des Particules - IN2P3, 74000 Annecy, France \label{LAPP} \and
IRFU, CEA, Universit\'e Paris-Saclay, F-91191 Gif-sur-Yvette, France \label{CEA} \and
Astronomical Observatory, The University of Warsaw, Al. Ujazdowskie 4, 00-478 Warsaw, Poland \label{UWarsaw} \and
Instytut Fizyki J\c{a}drowej PAN, ul. Radzikowskiego 152, 31-342 Krak{\'o}w, Poland \label{IFJPAN} \and
University of Oxford, Department of Physics, Denys Wilkinson Building, Keble Road, Oxford OX1 3RH, UK \label{Oxford} \and
Universit\'e Bordeaux, CNRS, LP2I Bordeaux, UMR 5797, F-33170 Gradignan, France \label{CENBG} \and
Institut f\"ur Physik und Astronomie, Universit\"at Potsdam,  Karl-Liebknecht-Strasse 24/25, D 14476 Potsdam, Germany \label{UP} \and
School of Physical Sciences, University of Adelaide, Adelaide 5005, Australia \label{Adelaide} \and
Friedrich-Alexander-Universit\"at Erlangen-N\"urnberg, Erlangen Centre for Astroparticle Physics, Erwin-Rommel-Str. 1, D 91058 Erlangen, Germany \label{ECAP} \and
Laboratoire Univers et Particules de Montpellier, Universit\'e Montpellier, CNRS/IN2P3,  CC 72, Place Eug\`ene Bataillon, F-34095 Montpellier Cedex 5, France \label{LUPM} \and
Institut f\"ur Astro- und Teilchenphysik, Leopold-Franzens-Universit\"at Innsbruck, A-6020 Innsbruck, Austria \label{Innsbruck} \and
Universit\"at Hamburg, Institut f\"ur Experimentalphysik, Luruper Chaussee 149, D 22761 Hamburg, Germany \label{UHH} \and
Obserwatorium Astronomiczne, Uniwersytet Jagiello{\'n}ski, ul. Orla 171, 30-244 Krak{\'o}w, Poland \label{UJK} \and
Institute of Astronomy, Faculty of Physics, Astronomy and Informatics, Nicolaus Copernicus University,  Grudziadzka 5, 87-100 Torun, Poland \label{NCUT} \and
Nicolaus Copernicus Astronomical Center, Polish Academy of Sciences, ul. Bartycka 18, 00-716 Warsaw, Poland \label{NCAC} \and
Institut f\"ur Astronomie und Astrophysik, Universit\"at T\"ubingen, Sand 1, D 72076 T\"ubingen, Germany \label{IAAT} \and
Institut f\"ur Physik, Humboldt-Universit\"at zu Berlin, Newtonstr. 15, D 12489 Berlin, Germany \label{HUB} \and
Department of Physics, University of the Free State,  PO Box 339, Bloemfontein 9300, South Africa \label{UFS} \and
Department of Physics and Astronomy, The University of Leicester, University Road, Leicester, LE1 7RH, United Kingdom \label{Leicester} 
\and
GRAPPA, Anton Pannekoek Institute for Astronomy, University of Amsterdam,  Science Park 904, 1098 XH Amsterdam, The Netherlands \label{Amsterdam} \and
Yerevan Physics Institute, 2 Alikhanian Brothers St., 375036 Yerevan, Armenia \label{Yerevan} \and
Kavli Institute for the Physics and Mathematics of the Universe (WPI), The University of Tokyo Institutes for Advanced Study (UTIAS), The University of Tokyo, 5-1-5 Kashiwa-no-Ha, Kashiwa, Chiba, 277-8583, Japan \label{KAVLI} \and
Department of Physics, Konan University, 8-9-1 Okamoto, Higashinada, Kobe, Hyogo 658-8501, Japan \label{Konan} \and
Department of Physics, Rikkyo University, 3-34-1 Nishi-Ikebukuro, Toshima-ku, Tokyo 171-8501, Japan \label{Rikkyo}
}
\offprints{H.E.S.S.~collaboration,
\protect\\\email{contact.hess@hess-experiment.eu};
\protect\\\protect\footnotemark[1] Corresponding authors
}

\abstract{Observations with imaging atmospheric Cherenkov telescopes (IACTs) have enhanced our knowledge of nearby supernova (SN) remnants with ages younger than 500 years by establishing Cassiopeia A and the remnant of Tycho's SN as very-high-energy (VHE) $\gamma$-ray sources. The remnant of Kepler's SN, which is the product of the most recent naked-eye SN in our Galaxy, is comparable in age to the other two, but is significantly more distant. If the $\gamma$-ray luminosities of the remnants of Tycho's and Kepler's SNe are similar, then the latter is expected to be one of the faintest $\gamma$-ray sources within reach of the current generation IACT arrays.

Here we report evidence at a statistical level of 4.6~$\sigma$ for a VHE signal from the remnant of Kepler's SN based on deep observations by the High Energy Stereoscopic System (H.E.S.S.) with an exposure of 152 hours. The measured integral flux above an energy of 226 GeV is $\sim$0.3\% of the flux of the Crab Nebula.
The spectral energy distribution (SED) reveals a $\gamma$-ray emitting component connecting the VHE emission 
observed with H.E.S.S. to the emission observed at GeV energies with \textit{Fermi}-LAT. The overall SED is similar to that of the remnant of Tycho's SN, possibly indicating the same nonthermal emission processes acting in both these young remnants of thermonuclear SNe.
}

\keywords{gamma-rays: general, supernovae: individual: Kepler's SN, ISM: supernova remnants, radiation mechanisms: non-thermal}

\maketitle

\makeatletter
\renewcommand*{\@fnsymbol}[1]{\ifcase#1\@arabic{#1}\fi}
\makeatother

\renewcommand{\labelitemi}{$\bullet$}

\section{Introduction}

For several decades, supernova remnants (SNRs) have been considered the most likely sources of Galactic 
cosmic rays 
\citep[CRs; e.g.,][]{ginzburgbook}, that is, CRs with energies at least up to $3\times 10^{15}$ eV. While the detection of radio and X-ray synchrotron radiation from SNRs does indeed prove that electrons are accelerated to GeV or even of order 10~TeV energies \citep[e.g.,][for reviews]{reynolds08,Helder2012,dubner15,vinkbook}, further insight into the particle acceleration in SNRs comes from $\gamma$-ray astronomy, which inter alia provides a probe of CR protons and nuclei through observations of GeV to TeV emission resulting from the decay of secondary neutral pions
produced in CR interactions.

Over the last two decades, SNRs have been established as an important population of Galactic $\gamma$-ray sources, both in the high-energy (HE) domain (100~MeV to 100~GeV) by the \textit{Fermi} Large Area Telescope (LAT) observations \citep{fermiSNRcat16} and in the very-high-energy (VHE) domain ($>100$~GeV) by observations 
with imaging atmospheric Cherenkov telescopes (IACTs); an SNR population study is provided in \citet[][]{hessSNRpop18} 
and an overview of VHE $\gamma$-ray astrophysics in \citet{hinton09}.

Very-high-energy spectra by themselves are often insufficient to firmly establish whether the $\gamma$-ray emission is produced by accelerated protons and other atomic nuclei
(so-called CR hadrons) or by accelerated electrons (CR leptons), or both. However, in some cases the presence of a characteristic pion-decay feature in the sub-GeV part of the $\gamma$-ray spectrum (the ``pion-decay bump'') can clearly establish a hadronic origin for the $\gamma$-ray emission. This is for example the case for SNRs IC 433
and W44 \citep{fermiPionBump}. In other cases, due to the fact that CR hadrons lead to 
a softer $\gamma$-ray spectrum (with a photon index of $\Gamma=2.0$ for a hadron power-law index of 2) in the HE regime 
than CR leptons 
($\Gamma=1.5$ in the Thomson regime for a lepton power-law index of 2), the spectral shape can indicate the production mechanism of the emission. However, a correspondence between spectral slopes and origin of $\gamma$-ray emission can be more complex if gas clumps are present \citep{gabici14} or in the case of nonlinear diffuse shock acceleration \citep{XXX1}, 
as well as for leptonic emission in the Klein-Nishina regime \citep{Porter2006}.

Supernova remnants with ages younger than 2000--3000~years have been established as TeV $\gamma$-ray sources.  
This fact suggests that CRs are accelerated to $>$10 TeV when the shock velocities are still in excess of a few 
thousand km~s$^{-1}$ \citep[][]{Helder2012}. Many of these young TeV $\gamma$-ray emitting SNRs belong to the class of historical SNRs, that is, SNRs produced by naked-eye supernovae (SNe) recorded in history, such as SN 185/RCW 86 \citep{rcw86hess18}, SN\,1006 \citep{sn1006}, and SN\,1572
\citep[also known as Tycho's SNR;][]{Archambault2017}. The young SNRs, RX J1713.7-3946 \citep[][]{refj1713} and particularly Cassiopeia A \citep[Cas A;][]{casa2017,casa2020}, which are the possible remnant of SN 393 \citep[][]{wang1997} and the remnant of a nearby 340-year-old SN missing in records, respectively, are also sometimes added to this SNR class. 
With these additions, both remnants of core-collapse SNe (RX J1713.7-3946 and Cas A) and 
thermonuclear (Type Ia) SNe (SN 185, SN\,1006, 
and SN\,1572/Tycho's  SNR) are included. Based on the $\gamma$-ray spectral characteristics, the $\gamma$-ray emission from Tycho's SNR and Cas A are likely of hadronic origin \citep{Morlino2012,casa2017,casa2020}, whereas for the other, more extended, 1000- to 3000-year-old SNRs a leptonic origin \citep[for RX J0852.0-4622; see][]{velaj1, velaj2, velaj3} has been suggested. However, a hadronic scenario for these older TeV \gray{} emitters has also been discussed, for example in \citet{Berezhko2012}, \citet{inoue12}, and \citet{gabici14}.

The remnant of the youngest naked-eye SN, SN 1604 \citep[also known as Kepler's SNR;][]{Green17,Vink2017}, 
is conspicuously absent
from the above list of young SNRs detected in VHE $\gamma$ rays. 
Kepler's SNR is, however, a known X-ray synchrotron emitter \citep{Allen1999, Nagayoshi2021}, indicating
that particles, at least electrons, are accelerated up to $\sim 10$~TeV. Moreover, the relatively
high density of its ambient medium \citep[e.g.,][]{Williams2012} makes Kepler's SNR a probable candidate for being a \gray{} source dominated by hadronic emission.
The High Energy Stereoscopic System (H.E.S.S.) experiment observed Kepler's SNR in the past, 
but the previous observations did not result in 
a detection of the remnant \citep{kepler2008}. This is partially due to the relatively short exposure 
time (13 hours) compared to other SNRs,
as well as the fact that Kepler's SNR, located at about 5 kpc \citep[][]{Ruiz2017}, is more distant than 
the historical SNRs listed above.

At GeV energies, an excess at about a 4$\sigma$ statistical level from Kepler's SNR was reported by \citet[][]{Xiang2021} and in the proceedings by \citet[][]{Prokhorov2021}. 
In \citet[][]{Xiang2021}, the authors claimed a detection of $\gamma$-ray emission from 
Kepler's SNR despite the presence of residual radiation around Kepler's SNR and 
concluded that more observation data with IACTs and \textit{Fermi}-LAT are necessary to firmly confirm the association between the $\gamma$-ray
source candidate and Kepler's SNR. Using a summed likelihood analysis and a larger
allowed range of zenith angles, \citet[][]{Acero2022} confirmed a detection of Kepler's SNR 
above a $6\sigma$ level based on 12 years of \textit{Fermi}-LAT data.

In this paper, we present the results of deep observations of Kepler's SNR performed with 
H.E.S.S. for a total of 152 hours. This data set is significantly larger than the
data comprising 13 hours of observations used in the previous publication \citep[][]{kepler2008}. 
We report first evidence for VHE $\gamma$-ray emission from Kepler's SNR - 
the last remaining historical SNR that had until now escaped detection at these energies. 
Combining the H.E.S.S. VHE data with HE data from 10.7 years of
\textit{Fermi}-LAT observations, we explore models to interpret the broad $\gamma$-ray 
spectral energy distribution (SED), in terms of hadronic and leptonic emission processes.

\section{Observations and analysis}

In this section, we present 
data reduction and analyses of H.E.S.S. and \textit{Fermi}-LAT data (sections 2.1.1 and
2.2.1), and results obtained from these analyses (sections 2.1.2 and 2.2.2).

\subsection{H.E.S.S. telescopes}

\begin{figure*}
\centering
  \begin{tabular}{@{}cc@{}}
    \includegraphics[angle=0, width=.48\textwidth]{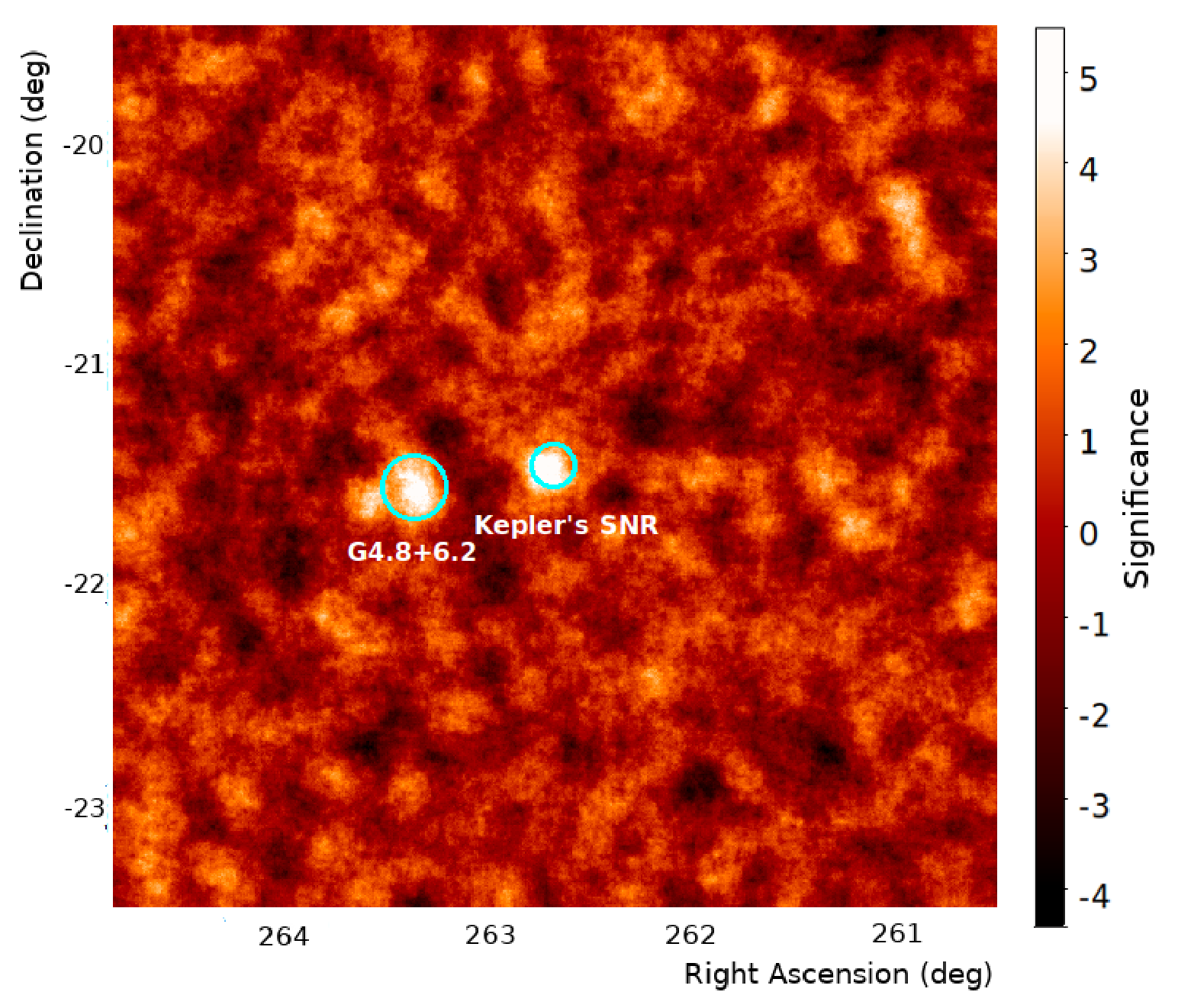}
    & 
    \includegraphics[angle=0, width=.48\textwidth]{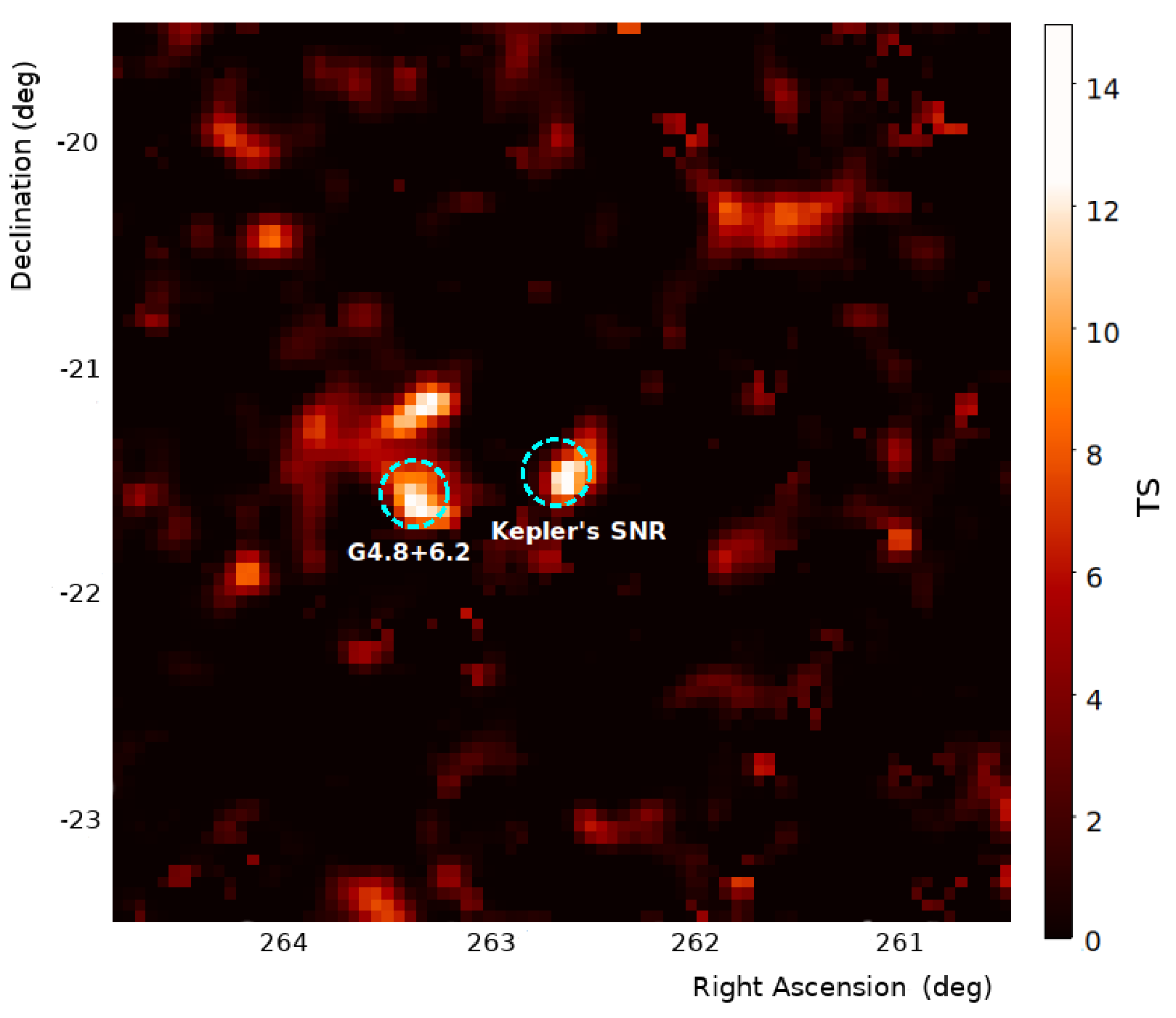} \\
   \end{tabular}
  \caption{H.E.S.S. and \textit{Fermi}-LAT maps. Left panel: H.E.S.S. \gray{} significance map
of Kepler's SNR using an oversampling radius of $0\fdg1$. Right
panel: \textit{Fermi}-LAT TS map in the range of 4.75-300 GeV.
The position of Kepler's SNR is in the center of these maps.
Solid circles in the left panel correspond to the source regions of Kepler's SNR  and 
SNR G4.8+6.2 (to the east of Kepler's SNR), while dashed circles in the right panel merely indicate the locations of these SNRs. Neither of these sources are present 
in the 4FGL catalog.} \label{F1}
\end{figure*}

The H.E.S.S. experiment is a hybrid array of five IACTs located in the southern hemisphere
in Namibia (23$^{\circ}$16$^{\prime}$18$^{\prime\prime}$ S,
16$^{\circ}$30$^{\prime}$00$^{\prime\prime}$ E) at an altitude of
$\sim$1800~m above sea level \citep[][]{Aharonian2006}. 
The H.E.S.S. array consists of four 12 m diameter 
telescopes \citep[][]{Ashton2020} placed in a square with 120 m sides and
one 28 m diameter telescope in the center of the array. 
A 28 m diameter telescope installed in 2012 complements the array, but is not 
used in this analysis.
Detection of VHE $\gamma$ rays with the four 12 m telescopes 
is based on the stereoscopic reconstruction technique.
We performed the dedicated observations of Kepler's SNR with H.E.S.S. reported 
in this paper using wobble mode with pointing offsets of $0\fdg7$ 
in right ascension or declination from Kepler's SNR, allowing a
simultaneous measurement of the background in the same field of view
\citep[][]{Berge2007}. 

\subsubsection{Data reduction and analysis}

We took data  
with at least three participating telescopes and in 28 min exposures called runs. 
We used the standard data quality selection procedure to identify observations with
satisfactory hardware states of the instrument and good atmospheric
conditions \citep[][]{Aharonian2006}. We analyzed 
152.2 h of good-quality data recorded during 353 runs. Of these, 122 h were accumulated in 2017-2020, whereas the 
remaining 30 h were spread over the years 2004-2013.
The mean zenith angle of these observations is
26$^{\circ}$. Given the age of Kepler's SNR, we assume its VHE \gray{} emission to be steady during the time interval of 16 years (4\% of the age of Kepler's SNR) in which the H.E.S.S. observations took place. This permits a simultaneous spectral fit of both the \textit{Fermi}-LAT and H.E.S.S. data. 

We analyzed the H.E.S.S. data using the \texttt{Model
Analysis} \citep[][]{Modelplus}
and used an analysis configuration that requires a minimum
of 60 photo-electrons per image and considers events with an estimated 
direction reconstruction uncertainty of less than $0\fdg1$. 
We cross-checked the results with the \texttt{Image Pixel-wise fit for Atmospheric
Cherenkov Telescope (\texttt{ImPACT})} analysis \citep[][]{ImPACT2014}.  
We defined a circular region-of-interest with
a radius of 6$^{\prime}$ around the geometrical center of
Kepler's SNR (at RA=17h30m40.8s,
Dec=-21$^{\circ}$29$^{\prime}$11$^{\prime\prime}$) - hereafter referred to as
the source region. This radius corresponds to the selection cut for a VHE point-like source. 
Kepler's SNR with $3.5^{\prime}$ diameter can be considered as a point-like source 
for H.E.S.S. enclosed within the source region.

\subsubsection{VHE $\gamma$-ray results}

The data analysis yields 1524 $\gamma$-ray-like
events from the source region and 23667 $\gamma$-ray-like events from a large ring-shaped background (off-source) region, 
from which 
we excluded the region with a $0\fdg3$ radius around another potential VHE \gray{} source (SNR G4.8+6.2; see Appendix A). 
The ratio of the on-source exposure to the off-source exposure, $\alpha$, is 0.0569.
We computed the number of excess counts by subtracting the number of off-source events multiplied
by $\alpha$ from the number of events coming from the source region. 
The \gray{} excess is 178 counts above the background, which corresponds to a significance of 
4.6$\sigma$ according to Eq. 17 from \citet[][]{LiMa1983}. The H.E.S.S. significance map is shown in the left panel of Fig. \ref{F1}. 
The main and cross-check analyses used in this paper provide compatible results.

We derived the energy spectrum 
using a reflected region background method \citep[][]{Berge2007} and a forward-folding technique \citep[][]{piron2001}. 
The analysis energy threshold for this data set is 226 GeV given by the energy at which 
the effective area falls to 15\% of its maximum value. The likelihood maximization yields a photon index of $\Gamma=2.3\pm0.2_{\mathrm{stat}}\pm0.2_{\mathrm{syst}}$ and a normalization constant of $N_{0}=(9.5\pm2.3_{\mathrm{stat}}\pm2.9_{\mathrm{sys}})\times10^{-14}$ cm$^{-2}$ s$^{-1}$ TeV$^{-1}$ at $E_{0}$=1 TeV
for a power-law spectrum $\mathrm{d}N/\mathrm{d}E=N_{0}\left(E/E_{0}\right)^{-\Gamma}$.
We binned the spectrum in such a way that spectral points require a minimum significance level of $2.0\sigma$ each.
The derived SED in the energy range from 226 GeV to 19 TeV is shown in Fig. \ref{F2}. The derived VHE flux is lower by a factor of 2.2 than the 99\% flux upper limit reported in the previous publication. The systematic uncertainties are conservatively estimated 
to be $\pm0.2$ on the photon index and $\pm30\%$ on the normalization coefficient \citep[][]{Aharonian2006}.

In addition to the VHE excess at the position of Kepler's SNR, Fig. \ref{F1} shows the presence
of a hotspot at the position of SNR G4.8+6.2 (see Appendix A), which is $\approx0\fdg7$ to the east
of Kepler's SNR. The analysis of this hotspot yields a $\gamma$-ray 
excess of 185 counts above the background. The numbers of on- and off-source events are $N_{\mathrm{ON}}=2007$ and
$N_{\mathrm{OFF}}=35049$, respectively, with a background-normalization factor $\alpha=0.0520$. 
This $\gamma$-ray excess is significant at a level of 4.2$\sigma$.

\subsection{\textit{Fermi}-LAT}

The principal instrument on board the \textit{Fermi Gamma Ray Space
Telescope} is the Large Area Telescope \citep[LAT;][]{Atwood2009}, a pair-conversion telescope
covering the energy range from about 20 MeV to more than 300 GeV.
\textit{Fermi}-LAT has been scanning the entire sky
continuously since August 2008.

\subsubsection{Data reduction and analysis}

As part of the motivation and preparation for the H.E.S.S. observations in 2020, 
we performed and took into account the analysis of \textit{Fermi}-LAT HE $\gamma$-ray 
data presented in this section. The presence of $\gamma$-ray excesses both in H.E.S.S. 
and \textit{Fermi}-LAT lends credence to the association between these $\gamma$-ray signals and Kepler's SNR.

For the data analysis, we used the \texttt{Fermitools} v1.2.23
package\footnote{\url{https://fermi.gsfc.nasa.gov/ssc/data/analysis/software/}}
and P8R3\_SOURCE\_V2 instrument response functions. We selected \textit{Fermi}-LAT
\gray{} events with reconstructed energies between 750 MeV and 300
GeV and performed a binned analysis by choosing a $10^{\circ}\times10^{\circ}$
square region centered at RA=17h30m48.7s,
Dec=-20$^{\circ}$1$^{\prime}$55$^{\prime\prime}$ (region of interest, ROI).
We shifted the position of the ROI center from that of Kepler SNR by $1\fdg4$ in
declination to reduce systematic effects due to the presence of a
molecular cloud, the Pipe Nebula, and the Galactic plane in the ROI
and their contributions to the foreground emission. 
For this analysis, we selected events accumulated from 2008 August 4 to 2019 May 16.
To reduce the contamination by the \gray{} emission
from the Earth's limb, we selected events with zenith angles
$<90^{\circ}$. We applied standard quality cuts
(\texttt{DATA}\_\texttt{QUAL>0} \&\&
\texttt{LAT}\_\texttt{CONFIG==1}). 

\begin{figure*}
\centering
  \begin{tabular}{@{}cc@{}}
    \includegraphics[angle=0, width=.48\textwidth]{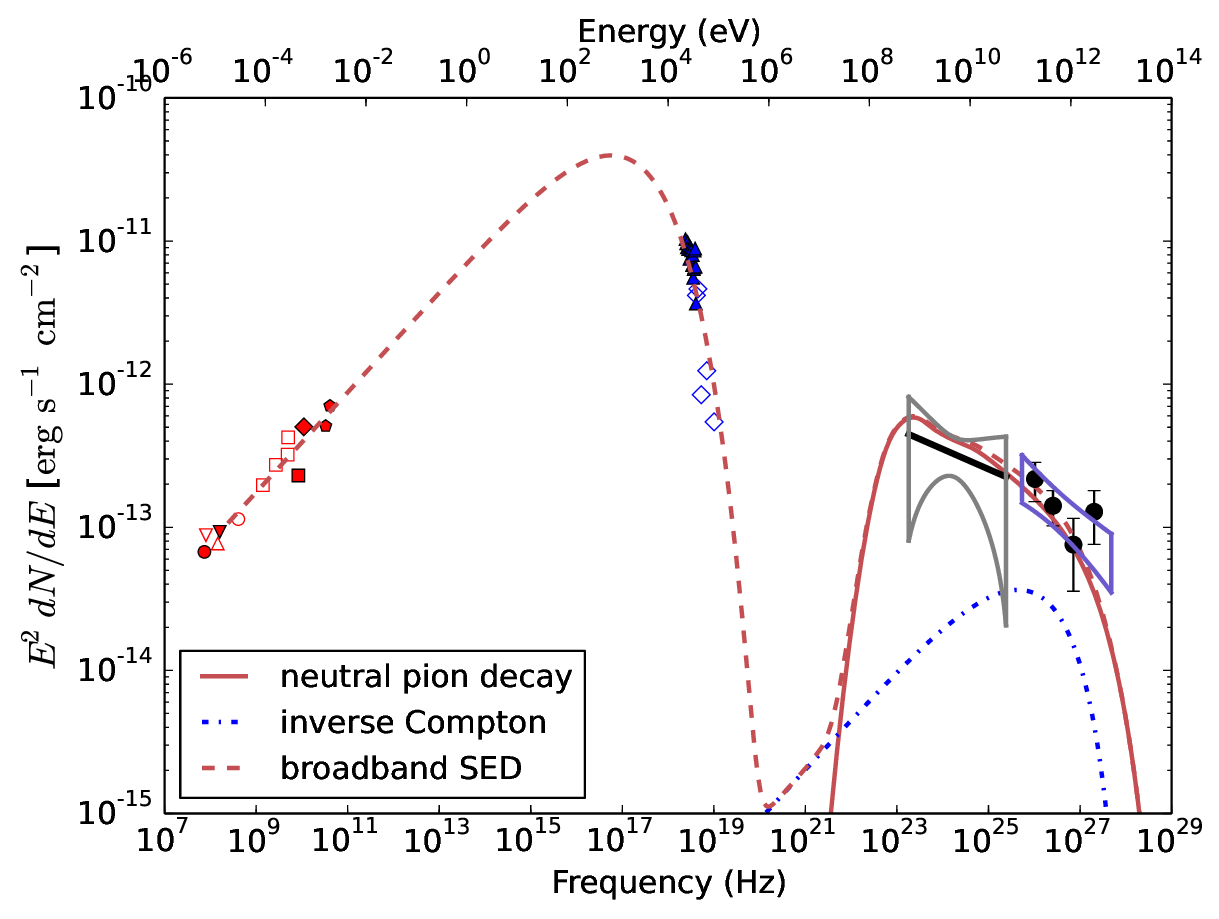}
    & \ \ \ 
    \includegraphics[angle=0, width=.48\textwidth]{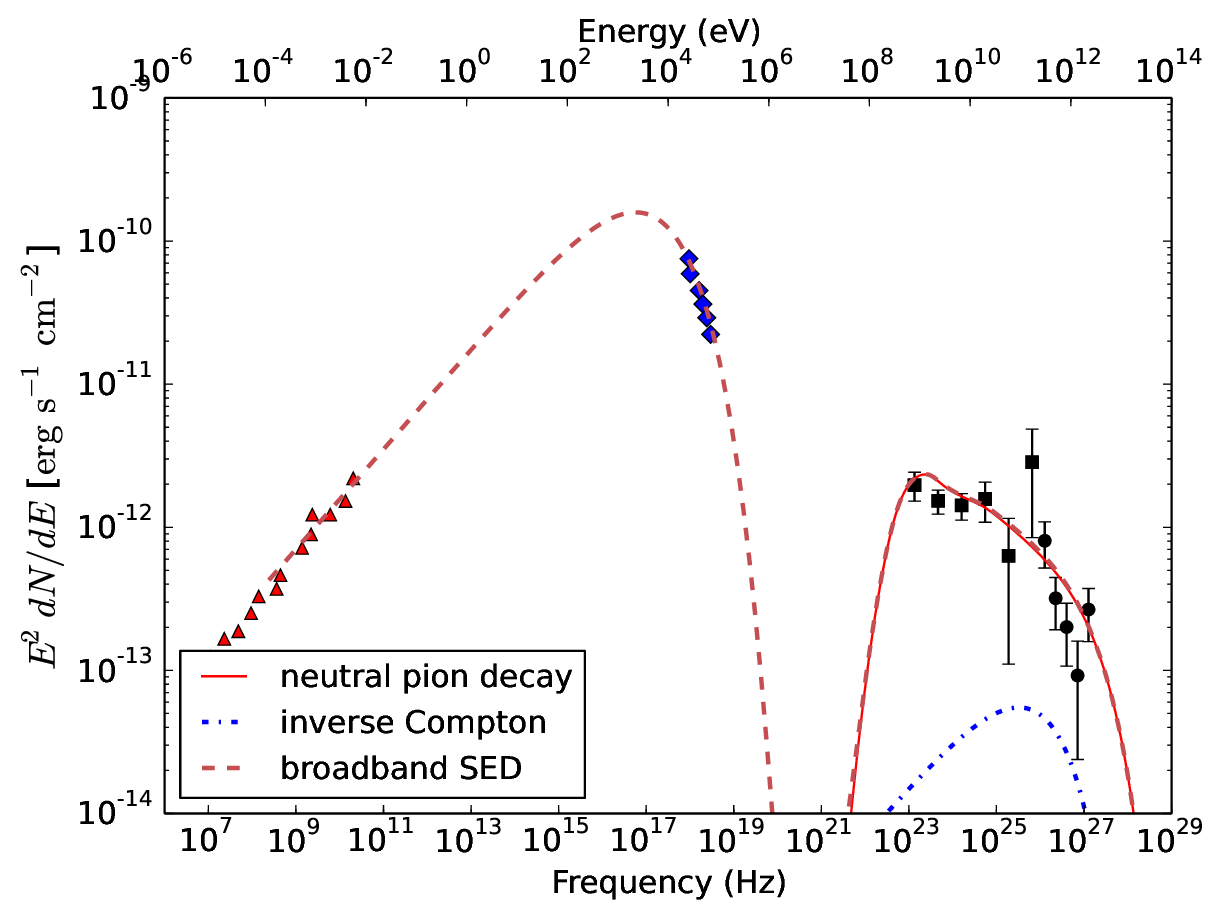}
   \end{tabular}
  \caption{Broadband SEDs of Kepler's SNR (left) and Tycho's SNR (right)
with hadronic model fits. SED data points and butterfly plots for Kepler's SNR 
as derived from H.E.S.S. and \textit{Fermi}-LAT data are shown. Observations of Kepler's SNR from the radio band to the X-ray band are from VLA (red filled $\circ$), Culgoora (red $\triangledown$), PAPER (red $\triangle$), Molonglo (red $\circ$), Parkes (red $\square$), NRAO (red $\diamond$), WMAP (red $\pentagon$), RXTE (blue $\triangle$),and Swift-BAT (blue $\diamond$).} \label{F2}
\end{figure*}

To model the sources within the ROI, we included sources from the
4FGL catalog \citep[][]{cat4fgl} within a region of
$17\fdg5$ radius around the center of the ROI. To model the
Galactic and isotropic background diffuse emission, we used the
standard templates \texttt{gll\_iem\_v07.fits} and
\texttt{iso\_P8R3\_SOURCE\_V2\_v1.txt}.  
The point spread function of \textit{Fermi}-LAT is sufficiently wide that Kepler's SNR 
can be treated as a point source.
We applied the binned
likelihood analysis on the data by using the \texttt{Fermitools} routine,
\texttt{gtlike}. We binned the data in 25 logarithmically spaced bands in energy and used the spatial binning with a pixel size of $0\fdg1$. We allowed the power-law normalization and photon index of Kepler's SNR
and the normalizations of the Galactic and isotropic diffuse sources to vary. We also allowed the normalizations of 4FGL \gray{} sources 
within $3^{\circ}$ of Kepler's SNR to vary, 
while keeping the normalizations of other 4FGL sources fixed.  
We evaluated the significance of model components or additional parameters using the
test statistic, TS = 2(log$\mathcal{L}$-log$\mathcal{L}_{0}$), where $\mathcal{L}_{0}$ is
the likelihood of the reference model without the additional
parameters or components \citep[][]{Mattox}. 

\subsubsection{HE $\gamma$-ray results}

For the 0.75-300 GeV energy interval, the derived TS value for Kepler's SNR is 16.8, which corresponds 
to about 4$\sigma$ significance. 
The likelihood maximization yields a photon index of
$\Gamma=2.1\pm0.3$ and a normalization constant,
$N_{0}=\left(9.3\pm3.1\right)\times10^{-15}$ cm$^{-2}$ s$^{-1}$
MeV$^{-1}$ at $E_{0}$=4750 MeV for a power-law spectrum. 
Figure \ref{F1} shows the TS map, based on a source model including
4FGL sources plus diffuse backgrounds, generated for the energy range of 4.75-300 GeV for the sake of comparison with the VHE $\gamma$-ray results. It reveals similarities between the \textit{Fermi}-LAT TS map and 
the H.E.S.S. significance map at VHE energies. 
The positive TS values and significances for Kepler’s SNR and SNR G4.8+6.2 
are present on both these maps in Fig. \ref{F1}. 
The butterfly shaped area shown in Figs. \ref{F2} and \ref{F3} corresponds to the 68\% statistical uncertainty region of the differential flux. 
The excesses both in \textit{Fermi}-LAT and H.E.S.S. data provide strong support for 
the identification of Kepler's SNR in the HE and VHE \gray{} bands.
Between 0.75-300 GeV, the residual emission at the position of SNR G4.8+6.2 is significant at a 3.3$\sigma$ 
statistical level (TS=10.9).


\section{Interpretation and modeling}
\label{sec4}

\subsection{Young SNRs as \gray{} sources}

The hadronic scenario---in which the \gray{} emission is produced
through the two-photon decay of neutral pions created in hadron collisions of
CRs with background gas---results in viable models for both the Tycho's and Cas A SNRs {\citep[][]{Zhang2013,casa2017}. 
The leptonic scenario, in which inverse Compton (IC) mechanism dominates the VHE emission, is still 
a viable scenario for Tycho's SNR under the assumption that its GeV \gray{}
emission is due to hadronic interactions \citep[][]{Yuan2012} or in the two-zone approach \citep{XXX2}.
However, more recently developed physical models suggest that the \gray{} emission
is primarily due to hadronic processes \citep[][]{Morlino2012,
Slane2014}.
Tycho's and Cas A SNRs are comparable in age with Kepler's SNR, which allows the radiative properties 
of SNRs to be investigated at an early stage in their evolution.

Both Type Ia and core-collapse SNe deliver similar kinetic energies
of $\sim 10^{51}$~erg. However, the type of SN does matter for both the ejected mass and the density and structure
of the ambient medium.
It is conceivable that the TeV
emission detected from the Cas A SNR, which resulted from a core-collapse explosion, originates dominantly from the reverse shock instead of
from the region heated by the forward shock \citep[e.g.,][]{helder08,Telezhinsky2013,Zirakashvili2014}. 
The reverse shock accelerates particles of the ejecta,
which, in turn, collide with the metal-rich plasma of the SN ejecta, leading to strong hadronic
emission due to the high ejecta density in core-collapse SNRs.
This circumstance might hinder a direct comparison of the radiation model for
the Cas A SNR with the radiation models for Tycho's and Kepler's
SNRs, for  which the \gray{} emission is thought to be dominated by forward-shock accelerated particles. 
The Galactic CR chemical composition, meanwhile, conforms better to 
acceleration of a well-mixed interstellar medium than metal-rich ejecta 
\citep[][]{Ellison1997}.
For thermonuclear SNe, more time is necessary for the progenitor system to evolve \citep[e.g.,][]{Tomonori};
this extra time allows it to wander a long distance
away from the parent molecular cloud.
Thus, the environment in which the remnants of thermonuclear SNe evolve can be significantly
different from the environment of core-collapse SNe of massive stars.

Multiwavelength studies of Kepler's SNR \citep[][for a review]{Vink2017} have  established that SN 1604
was a Type Ia SN. At a likely distance of 5~kpc \citep{sankrit16,Vink2017,Ruiz2017}, given the Galactic latitude of $6\fdg8$, Kepler's SNR is located 590~pc above
the Galactic plane. The relatively high density around Kepler's SNR, in particular in the northwestern part of the SNR, is attributed to mass lost by the progenitor
system \citep{bandiera87,chiotellis12,Williams2012,Burkey2013,toledo14}.
Apart from its relatively large distance and high Galactic latitude,
Kepler's SNR is best compared to Tycho's SNR located at a distance of about 2.5~kpc: both are Type Ia
SNRs of similar age and develop in a relatively dense ambient medium with $n\approx 1$--10~cm$^{-3}$.
}

\subsection{SED models}

A variety of multiwavelength data is available for Kepler's SNR.
It allows us to construct a characteristic SED for the nonthermal radiation caused by
accelerated particles. 
In addition to the $\gamma$-ray data points described above,
we use radio and X-ray spectral data points of Kepler's SNR
from \citet[][]{Reynolds1992, Allen1999} along with the WMAP
data points at 33 and 41 GHz from \citet[][]{Massardi2009} and the
\textit{Swift}-BAT data points at (4.2, 4.5, 5.3, 7.0, and
10.0)$\times10^{18}$ Hz from \citet[][]{Oh2018}. 
For comparison with the SED of
Tycho's SNR, the data points in the radio and X-ray bands are 
from \citet[][and references therein]{Zhang2013}, and the \textit{Fermi}-LAT and VERITAS \gray{}
data points are from \citet[][]{Archambault2017}.
 The properties of Kepler's and 
Tycho's SNRs are described in the Appendix B.

In the framework of a hadronic model, we chose typical numerical values for physical parameters and
described both SEDs of Kepler's and Tycho's SNRs by using the modeling
package, \texttt{Naima} \citep[][]{Zabalza2015}. 
The radio band to X-ray band SEDs are attributed to the synchrotron radiation 
of a CR electron population, while the broad $\gamma$-ray SED is attributed to 
radiation driven by the collisions of relativistic CR protons with
gas.
Our model setup includes:
(i) $E_{\mathrm{sn}}=10^{51}$ erg for the SN Ia explosion energy;
(ii) the distances to Kepler's and Tycho's SNRs of 5.0 kpc
\citep[][]{Ruiz2017} and 2.5 kpc \citep[][]{Zhang2013,
Kozlova18}, respectively; (iii) $n=1$ cm$^{-3}$ for the gas target
particle density; (iv) the kinetic energy of CR hadrons above 1 GeV of $E_{\rm CR}/E_{\rm SN}=$7\% of the SN Ia
explosion energy; (v) $B=200$ $\mu$G for the magnetic field
strength\footnote{The arbitrary value of 200 $\mu$G is chosen so as to suppress the IC-to-synchrotron emission ratio and 
to remain within reasonable downstream magnetic field ranges for these remnants.} \citep[][]{Voelk2005, Helder2012}; 
(vi) the ratio of energy in CR electrons to that in CR
hadrons is $0.5$\%, constrained by fitting the
synchrotron SED component; (vii) $q=2.3$ for the CR 
electron and proton spectral indices, $dN_{\mathrm{p, e}}/dE\propto E^{-q}$,
assumed to be the same and chosen to provide us with a fit in the 
radio-frequency band \citep[][]{DeLaney2002, Kothes2006};
and (viii) exponential cut-off energies in the CR proton spectrum at 100 TeV and 
in the CR electron spectrum at 7.5 TeV, respectively. 
We assumed a fraction of the SN explosion energy converted into the energy of CRs similar to that given by \citet[][]{berezhko2006} (that is 10\%). 
The parameter values (iii), (vi), and (viii) 
are chosen such that the hadronic model matches the SEDs of Kepler's and Tycho's SNRs. 
The constraint on the proton exponential cut-off energy is however premature, since
the power-law hypothesis is currently sufficient to describe the observed SED in the VHE band.
We note that the model is sensitive to the combination $E_{\rm CR}\times n$ 
rather than these two parameters individually.
A model-dependent constraint on the density can be set by the size of the SNR at its given age during the adiabatic (Sedov) stage \citep[e.g.,][]{berezhko2006}.

The left and right panels in Fig. \ref{F2} show the broadband SEDs of Kepler's and 
Tycho's SNRs, respectively, along with the hadronic model described above.
The observations support the similarity between the shapes of their broadband SEDs.
In the framework of this model,
the hadronic \gray{} component dominates over the \gray{} component
produced by CR electrons.

A comparison of the results presented in Fig. \ref{F2} shows that the same
hadronic model scaled according to the different distances to the two SNRs describes both the SEDs of Kepler's and Tycho's SNRs. 
This conclusion supports that these remnants, similar in at least SN type and age, have also similar broadband nonthermal emission properties.
The luminosities of Tycho's and Kepler's SNRs at 1 TeV are similar for the assumed distances, whereas the TeV luminosity of Cas A is five times higher. However, the TeV luminosity depends on several parameters, such as the CR energy in the source and the average density. So there is a possibility that the similarity of Tycho's and Kepler's SNR may hide some underlying differences in these parameters.
This SED model can also describe the SEDs of Kepler's and Tycho's SNRs
even if they are located at 7 kpc and 3.5 kpc, respectively, keeping the same distance ratio as above (that is 5/2.5=2),
but increasing the gas density from $1$ cm$^{-3}$ to $(7/5)^2\times1$ cm$^{-3}$ $=2$ cm$^{-3}$.
The gas density around Kepler's SNR has a strong gradient, with the
southeastern part being more tenuous than the northwestern part. Moreover, the gas
appears to be clumpy. Hence, gas densities vary from $n\sim 1$--250~cm$^{-3}$ \citep{Williams2012}, and the average gas density is not well constrained.
The distribution of $\gamma$-ray emission from different gas accumulations in these remnants remains hidden in the model. In future works on modeling, questions to be addressed are the implementation 
of a gas density distribution, the efficiency of conversion of kinetic energy to HE particles,
the cutoff shape in the electron spectrum, and the connection to the dynamical
evolution of the SNRs.
Future \gray{} studies of Kepler's SNR are required for precise determination of its spectral and morphological 
characteristics in the VHE \gray{} band and further examination of this joint interpretation.

In a leptonic scenario, the \gray{} emission is produced via IC scattering 
of photons from the cosmic microwave background (CMB), the infrared photon
field emitted by dust in SNRs, and the Galactic interstellar radiation 
field by energetic electrons. To compute the IC \gray{} component, we included these three
photon fields making similar contributions and took energy densities and spectra of the latter
two photon fields from \citet[][]{Gomez2012} and
\citet[][]{Porter2006}. 
Since the same electron population is emitting both X-rays via the synchrotron mechanism and the HE and VHE 
\grays{} via the IC mechanism, we constrain
the magnetic-field strength by requiring that the model reproduces 
the observed flux of X-ray emission and does not overshoot the H.E.S.S. 
SED data points. 
In case the $\gamma$-ray emission is dominated by hadronic processes, this constraint
should be regarded as a lower limit to the magnetic-field strength \citep[][]{kepler2008}.
From the \gray{} observations reported above, it follows that 
for magnetic field values
greater than 80 $\mu$G the resulting IC flux would be less than the measured $\gamma$-ray flux. 

\begin{figure}
\centering
    \includegraphics[angle=0, width=.48\textwidth]{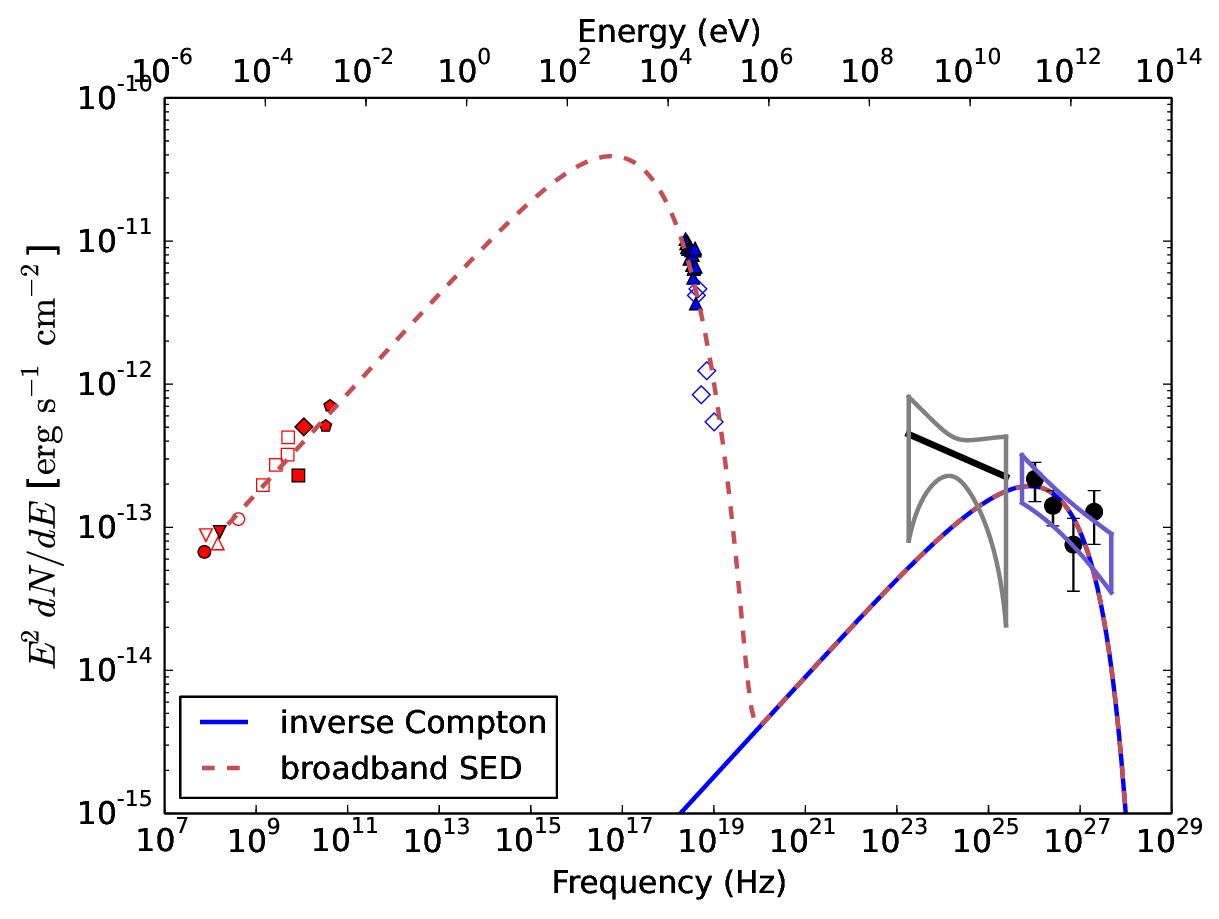}
  \caption{Broadband SED of Kepler's SNR with a leptonic model fit corresponding to the magnetic field of 80 $\mu$G.}
\label{F3}
\end{figure}

Figure~\ref{F3} illustrates the leptonic model with a magnetic field strength of 
80 $\mu$G. 
This magnetic-field strength is marginally below than the measured downstream 
magnetic field strength value for Kepler's SNR, which is in the range of 
100-200$\mu$G \citep[][]{Voelk2005, Helder2012}. 
The photon index measured from the \textit{Fermi}-LAT data, $\Gamma= 2.1\pm0.3$, is softer 
than that expected for the leptonic scenario, $\Gamma=1.7$.
In order to better determine which emission scenario is preferred,
we compared the two models using the likelihood ratio test based on the \textit{Fermi}-LAT data 
with energies between 750 MeV and 300 GeV. 
The null model assumes emission via IC scattering with a $\gamma$-ray spectrum fit to the H.E.S.S. data.
Between 750 MeV and 300 GeV, the $\gamma$-ray spectrum in the null model is well approximated by
a power law with fixed parameters.
The alternative model assumes a power-law model with a free flux normalization and a free photon index,
and the best-fit spectral parameters in this model better correspond to the hadronic scenario. 
The likelihood analysis described in Section 2.2 gives the TS
value 5.1 with 2 degrees of freedom for a comparison of the null and alternative models. 
Therefore, the spectral shape favors a hadronic model over a leptonic model with a 1.8$\sigma$ significance. \citet[][]{Acero2022} have recently performed an analysis of 12 years
of the \textit{Fermi}-LAT data, selecting $\gamma$ rays above 100 MeV. The 
derived photon index value is in agreement with that reported above and their lowest energy
flux point provides further support for a hadronic model.
The SED at $\gamma$-ray energies can also be explained by a mixed scenario involving both leptonic and hadronic contributions.

\section{Summary and conclusions}

Gamma-ray observations of young SNRs provide insights into the
particle acceleration mechanisms and HE emission processes. 
HE and VHE \gray{} observations of the Tycho and
Cas A SNRs show that remnants younger than 500 years old emit $\gamma$ rays 
likely via the decay of neutral pions created
in proton-proton collisions. Sharing a number of the physical
properties with Tycho's SNR, Kepler's SNR is an important testbed for
a connection between CRs and SNRs.

This paper reports the results of 152 hours of VHE observations
of Kepler's SNR with H.E.S.S. and provides a characterization of
the whole data set using advanced analysis methods. 
This data set is significantly larger than that used in the previous publication.
VHE \gray{} emission from Kepler's SNR is significant at a $4.6\sigma$ statistical level. A spectral analysis of this data set yields a photon index of $\Gamma=2.3\pm0.2_{\mathrm{stat}}\pm0.2_{\mathrm{sys}}$ and a normalization constant of $N_{0}=(9.5\pm2.3_{\mathrm{stat}}\pm2.9_{\mathrm{sys}})\times10^{-14}$~cm$^{-2}$~s$^{-1}$~TeV$^{-1}$ at $E_{0}=1$~TeV.
This paper also includes an analysis of 10.7 years of \textit{Fermi}-LAT data performed 
before making the decision on H.E.S.S. observations of Kepler's SNR in 2020. This analysis 
provides an indication of a GeV counterpart of Kepler's SNR and confirms the results reported 
by \citet[][]{Xiang2021} and \citet[][]{Acero2022}. 
The compatibility of the signals in the HE and VHE \gray{} bands supports their common origin 
and association with Kepler's SNR. 

Although the detection at the 4.6$\sigma$ level falls short of the
gold-standard in the field of VHE $\gamma$-ray astronomy of $>5\sigma$, 
our results are based on observations targeting Kepler's SNR directly, and not a blind search. Moreover, Kepler's SNR, being a young SNR and an X-ray synchrotron source, is a priori expected to be a VHE \gray{} source.
These arguments altogether provide a strong reason to
endorse Kepler's SNR as a faint HE and VHE $\gamma$-ray source.

The results show that the observed SED of Kepler's SNR is similar to that of Tycho's SNR, 
possibly indicating the same nonthermal emission processes acting in both these SNRs. 
This fact allows us for the first time to tentatively propose a model
of \gray{} emission that describes both the SEDs of Kepler's and Tycho's
SNRs. It assumes a hadronic origin of the observed $\gamma$-ray emission 
and requires (1) $\approx10$\% of the SN Ia explosion energy to be converted
to  CR hadron energy and (2) the gas target particle density of $\approx1$
cm$^{-3}$. 
A lower limit on the magnetic field strength of $B>80$ $\mu$G
derived from the requirement that the IC \gray{} component does not
overshoot the H.E.S.S. SED data points is tighter than the previous lower limit
\citep[][]{kepler2008} by a factor of $\approx$1.5.

\begin{acknowledgements}
The support of the Namibian authorities and of the University of Namibia in facilitating the construction and operation of H.E.S.S. is gratefully acknowledged, as is the support by the German Ministry for Education and Research (BMBF), the Max Planck Society, the German Research Foundation (DFG), the Helmholtz Association, the Alexander von Humboldt Foundation, the French Ministry of Higher Education, Research and Innovation, the Centre National de la Recherche Scientifique (CNRS/IN2P3 and CNRS/INSU), the Commissariat à l'énergie atomique et aux énergies alternatives (CEA), the U.K. Science and Technology Facilities Council (STFC), the Knut and Alice Wallenberg Foundation, the National Science Centre, Poland grant no.2016/22/M/ST9/00382, the South African Department of Science and Technology and National Research Foundation, the University of Namibia, the National Commission on Research, Science $\&$ Technology of Namibia (NCRST), the Austrian Federal Ministry of Education, Science and Research and the Austrian Science Fund (FWF), the Australian Research Council (ARC), the Japan Society for the Promotion of Science and by the University of Amsterdam. 
We appreciate the excellent work of the technical support staff in Berlin, Zeuthen, Heidelberg, Palaiseau, Paris, Saclay, Tübingen, and in Namibia in the construction and operation of the equipment. This work benefitted from services provided by the H.E.S.S. Virtual Organisation, supported by the national resource providers of the EGI Federation. 
J. Vink and D. Prokhorov are partially supported by funding from the European Union’s Horizon 2020 research and innovation programme under grant agreement No 101004131 and by the Netherlands Research School for Astronomy (NOVA).
\end{acknowledgements}

\bibliographystyle{aa}

\appendix

\section{SNR G4.8+6.2}

In this appendix, we describe the properties of a $\gamma$-ray source candidate, SNR G4.8+6.2.

SNR G4.8+6.2 is located approximately 40 arcmin away from Kepler's SNR.
The physical properties of SNR G4.8+6.2 are not well known. In the radio band, this SNR has a shell-like morphology and an angular extent of 18$^{\prime}$ at 1.4 GHz (the NRAO VLA Sky Survey). At 2.3 GHz it appears highly polarized with an almost constant orientation of the polarization vectors across the source and with the mean fraction of polarized emission of up to 25\% \citep[][]{Duncan1997}. Young SNRs, such as Kepler's SNR, have a much smaller fractional polarization. 

\section{Tycho's and Kepler's SNRs}

In this appendix, we describe the properties of Tycho's and Kepler's SNRs.

The distance to Tycho's SNR is estimated to be 2.5 kpc \citep[2.5 kpc or $2.8\pm0.4$ kpc from][respectively]{Zhang2013, Kozlova18}, while the distance to Kepler's SNR is $5.0\pm0.7$ kpc \citep[][]{Ruiz2017}. Their measured forward shock velocities are in excess
of $\sim2000$ km s$^{-1}$ (see, for example, \citealt{Hwang02, Williams2016} for Tycho's
SNR, and \citealt{Vink2008, Coffin2022} for Kepler's SNR). The downstream
magnetic field strength for these SNRs are in the range of
100-200$\mu$G \citep[][]{Voelk2005, Helder2012}. The observed
expansion rates of both Tycho's and Kepler's SNRs indicate that
these SNRs are still in transition from the early expansion phase to
the Sedov phase \citep[][]{Reynoso97, Hughes00, Vink2008}. The X-ray
spectra of both these SNRs with the presence of large amounts of
iron \citep[e.g.,][]{Hwang1998, Kinugasa99} show that they are
almost certainly of type Ia, and their classification is proven by
the spectrum of a scattered-light echo from Tycho's SN
\citep[][]{Krause08, Rest08} and supported by the location of Kepler
SNR well above the Galactic plane, respectively. Tycho's SNR, located
at a Galactic latitude of $1\fdg4$, is probably
interacting with a semi-closed large molecular shell surrounding the
SNR from the north to the east \citep[][]{Ishihara2010, Zhang2013},
while Kepler's SNR, located in a lower density medium, has a density
gradient with densities in the north greater by an order of
magnitude than those in the south \citep[][]{Williams2012}. 
The centroid position of the $\gamma$-ray emission 
from Tycho's SNR is consistent with the center of the shell rather 
than offset toward the northeastern region \citep{Archambault2017}.
For both Tycho's and Kepler's SNe a single degenerate channel
seems likely, given the presence of dense circumstellar gas with
which the shock waves are interacting.
For Kepler's SNR this is even more remarkable given its height above the Galactic
plane, making a chance coincidence highly unlikely. 
Inside Tycho's SNR, a possible surviving companion star, known as Tycho-G, has been identified
\citep[][]{Ruiz2004, Bedin14}, although the identification has been disputed \citep{Kerzendorf2013}.
Inside Kepler's SNR no surviving donor star has been found, which is difficult to reconcile
with a single degenerate channel, but could perhaps suggest that the donor star
evolved into a (proto) white dwarf not long before the explosion. This is the so-called
core-degenerate channel \citep{Ilkov2012,Vink2017}.
The delayed-detonation model for
Type Ia SNe \citep[][]{Khokhlov1991} is the most probable explosion
mechanism for both Tycho's and Kepler's SNe
\citep[e.g.,][]{Badenes2003, Sun2019}, resulting in the explosion
energy of $E_{\mathrm{SN}}=(1.3-1.6)\times10^{51}$ erg
\citep[][]{Gamezo2005}. The light curves of Tycho's and 
Kepler's SNe 
suggest that these events are SNe Ia 
with the normal rate of decline after maximum light \citep[][]{Ruiz2004, Vink2017, Ruiz2017}, but not slowly declining, 
overluminous SNe Ia (such as SN 1991T) or fast declining, underluminous SNe Ia (such as SN 1991bg).
On the other hand, X-ray line intensity ratios of iron-group elements to intermediate-mass elements for Kepler's SNR are 
much higher than that for Tycho's SNR and correspond better to
an overluminous event \citep[][]{Katsuda2015}, and the amount of iron emission
in the X-ray spectrum of Kepler's SNR rules out the subenergetic models \citep[]{Patnaude2012}.

\end{document}